\documentclass{elsarticle}
\usepackage[utf8]{inputenc}
\usepackage[caption=false]{subfig}
\usepackage{multicol}
\usepackage{multirow}
\usepackage{booktabs}
\usepackage{amsmath}
\usepackage[normalem]{ulem}

\usepackage{xcolor}
\usepackage[colorinlistoftodos]{todonotes}
\usepackage{array}
\newcolumntype{P}[1]{>{\centering\arraybackslash}p{#1}}
\newcolumntype{M}[1]{>{\centering\arraybackslash}m{#1}}

\title{Anisotropic 3D Multi-Stream CNN for Accurate Prostate Segmentation from Multi-Planar MRI}
\author[1]{Anneke Meyer\corref{cor1}\fnref{fn1}}

\author[2,3]{Grzegorz Chlebus\fnref{fn1}}
\author[1]{Marko Rak}
\author[4]{Daniel Schindele}
\author[4]{Martin Schostak}
\author[3,2]{Bram van Ginneken}
\author[2]{Andrea Schenk}
\author[5,2]{Hans Meine}
\author[2]{Horst K. Hahn}
\author[2]{Andreas Schreiber}
\author[1]{Christian Hansen}

\cortext[cor1]{Corresponding author. \newline
\textit{Email-Address:} anneke.meyer@ovgu.de. \newline
    \textit{Postal Address: } Universitaetsplatz 2, 39106 Magdeburg, Germany} 
\fntext[fn1]{Anneke Meyer and Grzegorz Chlebus contributed equally to this work.\\
\\
\copyright\,2020. This manuscript version is made available under the CC-BY-NC-ND 4.0 license http://creativecommons.org/licenses/by-nc-nd/4.0/.}

\address[1]{Faculty of Computer Science and Research Campus STIMULATE, University of Magdeburg, Germany}
\address[2]{Fraunhofer Institute for Digital Medicine MEVIS, Bremen, Germany}
\address[3]{Radboud University Medical Center, Nijmegen, The Netherlands}
\address[4]{Clinic of Urology and Pediatric Urology, University Hospital Magdeburg, Germany}
\address[5]{University of Bremen, Medical Image Computing Group, Bremen, Germany}

\date{November 2020}
\journal{Elsevier Comput Methods Programs Biomed}
\usepackage{etoolbox}
\makeatletter
\patchcmd{\ps@pprintTitle}
  {Preprint submitted to}
  {Accepted manuscript in }
  {}{}
\makeatother

\begin{document}



\begin{abstract}
    \textit{Background and Objective:} Accurate and reliable segmentation of the prostate gland in MR images can support the clinical assessment of prostate cancer, as well as the planning and monitoring of focal and loco-regional therapeutic interventions. Despite the availability of multi-planar MR scans due to standardized protocols, the majority of segmentation approaches presented in the literature consider the axial scans only. In this work, we investigate whether a neural network processing anisotropic multi-planar images could work in the context of a semantic segmentation task, and if so, how this additional information would improve the segmentation quality. \\
    \textit{Methods:} We propose an anisotropic 3D multi-stream CNN architecture, which processes additional scan directions to produce a high-resolution isotropic prostate segmentation. We investigate two variants of our architecture, which work on two (dual-plane) and three (triple-plane) image orientations, respectively. The influence of additional information used by these models is evaluated by comparing them with a single-plane baseline processing only axial images. To realize a fair comparison, we employ a hyperparameter optimization strategy to select optimal configurations for the individual approaches.\\
    \textit{Results:} Training and evaluation on two datasets spanning multiple sites show statistical significant improvement over the plain axial segmentation ($p<0.05$ on the Dice similarity coefficient). The improvement can be observed especially at the base ($0.898$ single-plane vs. $0.906$ triple-plane) and apex ($0.888$ single-plane vs. $0.901$ dual-plane).\\
    \textit{Conclusion:} This study indicates that models employing two or three scan directions are superior to plain axial segmentation. The knowledge of precise boundaries of the prostate is crucial for the conservation of risk structures. Thus, the proposed models have the potential to improve the outcome of prostate cancer diagnosis and therapies.
\end{abstract}


\maketitle

\section{Introduction}

Prostate cancer is the most prevalent type of cancer among men accounting for over 164 thousand new cases and more than 29 thousand deaths in the US in 2018~\cite{siegel2018cancer}.
Clinical workflows of prostate cancer patients commonly involve MR imaging, which, thanks to the high soft-tissue contrast, can be employed for diagnosis, staging, and therapy planning.
Prostate segmentation in MRI is a time-consuming task, requiring expert knowledge and suffering from inter-observer variability.
Knowledge of the gland size and shape, which can be derived from the segmentation mask, is often utilized in clinical and research applications.
For instance, Shah \textit{et al.}\,\cite{Shah2009-ly} has shown that MRI findings can be correlated with the prostatectomy specimen by employing the prostate segmentation.
Moreover, it is often used to facilitate radiotherapy planning~\cite{Schmidt2015radiotherapy} and targeted biopsy with MRI-TRUS (transrectal ultrasound) fusion \cite{Fedorov2015-wn, Das2020}.
Because neighboring structures as seminal vesicles, bladder, neurovascular bundles, and the external sphincter are essential for the erectile function and urine continence of men, the segmentation should be as precise as possible for the planning of prostate cancer therapy.

\subsection{Related Work}
Before the advance of deep learning, prostate segmentation was mainly performed with atlas-based segmentation or deformable models based on hand-crafted features. A comprehensive summary of those methods is given in \cite{Ghose2012-pp}. Early approaches incorporating deep learning used voxel-wise classification to yield a segmentation mask. For instance, Liao \textit{et al.} \cite{Liao2013-nm} learned deep features with a stacked independent subspace analysis network in an unsupervised fashion and perform segmentation with label propagation from atlases. Guo \textit{et al.} \cite{Guo2016-ua} also used deep features but generated by a supervised stacked sparse autoencoder, yielding a prostate likelihood map, which is then segmented by a deformable model. Jia \textit{et al.} \cite{jia2017prostate} performed patch-based prediction with ensemble deep convolutional neural networks (CNNs).

CNNs are gaining attention in the medical image processing field thanks to state-of-the-art results on numerous classification and segmentation tasks.
Various CNN architectures for segmentation problems were proposed.
Long \textit{et al.} proposed a fully convolutional neural network (FCN), which can be applied to arbitrarily sized images~\cite{long2015fully}.
The U-Net model by Ronneberger \textit{et al.} following the encoder/decoder design with long skip connections to retain the locality information was successfully used for different image segmentation problems~\cite{ronneberger2015u}.
Established CNN architectures, as well as their modified versions, have been introduced for prostate segmentation on T2-weighted MRI.
For instance, Tian~\textit{et al.} fined-tuned a FCN model for prostate segmentation~\cite{Tian2017-pv}.
Yan \textit{et al.} \cite{yan2019propagation} adopted a FCN to embed superpixel information as low-level features in combination with high-level deep features. Another modification strategy to improve network segmentation is to add deep supervision\,\cite{Zhu2017-mt, cheng2017automatic, Wang.2019, Wang.16.02.201921.02.2019}.

Learning and segmentation performance can benefit from different aspects regarding network design to retain fine-detailed information and alleviate the vanishing gradient problem. While the U-Net architecture employs skip connections from the encoder to the decoder part of the network, Yu \textit{et al.} \cite{yu2017volumetric} analyzed the effect of short and long residual connections and showed that a combination of both is beneficial in a 3D~CNN for segmentation. Wang \textit{et al. }\cite{Wang.2019} observed improvements with residual connections between neighboring blocks in combination with strided convolutions. Hossain \textit{et al.}\,\cite{Hossain.2018} adapted the VGG19 architecture \cite{Simonyan.942014} into an FCN and added short and long residual connections. A ResNet \cite{He.2016} encoder was extended with a decoder with 3D global convolutional block and boundary refinement blocks in \cite{Jia.18.07.2018}. The authors combined this network with an adversarial network for higher-order consistent predictions. In the whole model, anisotropic convolutions are employed to reflect the high slice thickness of the MR input volumes. The authors furthermore suggested using the ResNet encoder in combination with an anisotropic decoder and multi-level pyramid convolutional skip connections as well as adversarial training \cite{Jia.2019}.

The use of dense connections that enhance feature reuse and propagation has been shown in the last two years to improve performance additionally. Hassanzadeh \textit{et al.} \cite{Hassanzadeh.2019} evaluated the use of various residual and dense connections. Yuan \textit{et al.} \cite{Yuan.2019} made use of densely connected blocks in encoder and decoder and trained with a joint loss function that incorporates the Dice similarity coefficient and the reconstruction error of dense block outputs. Also, Zhu \textit{et al.} \cite{Zhu.21.02.2019, Zhu.2018b}, To \textit{et al.} \cite{To.2018} and Liu \textit{et al. }\cite{Liu.1312018} incorporated, amongst others, dense blocks into their architectures. Brosch \textit{et al.} \cite{Brosch.2018} formulated the segmentation as a regression task. They combined a 3D shape model with a convolutional regression network, where the network is used to obtain the distance from the surface mesh to the corresponding boundary point of the prostate.

\begin{figure}[!t]
\centering
\centerline{\includegraphics[width =\textwidth]{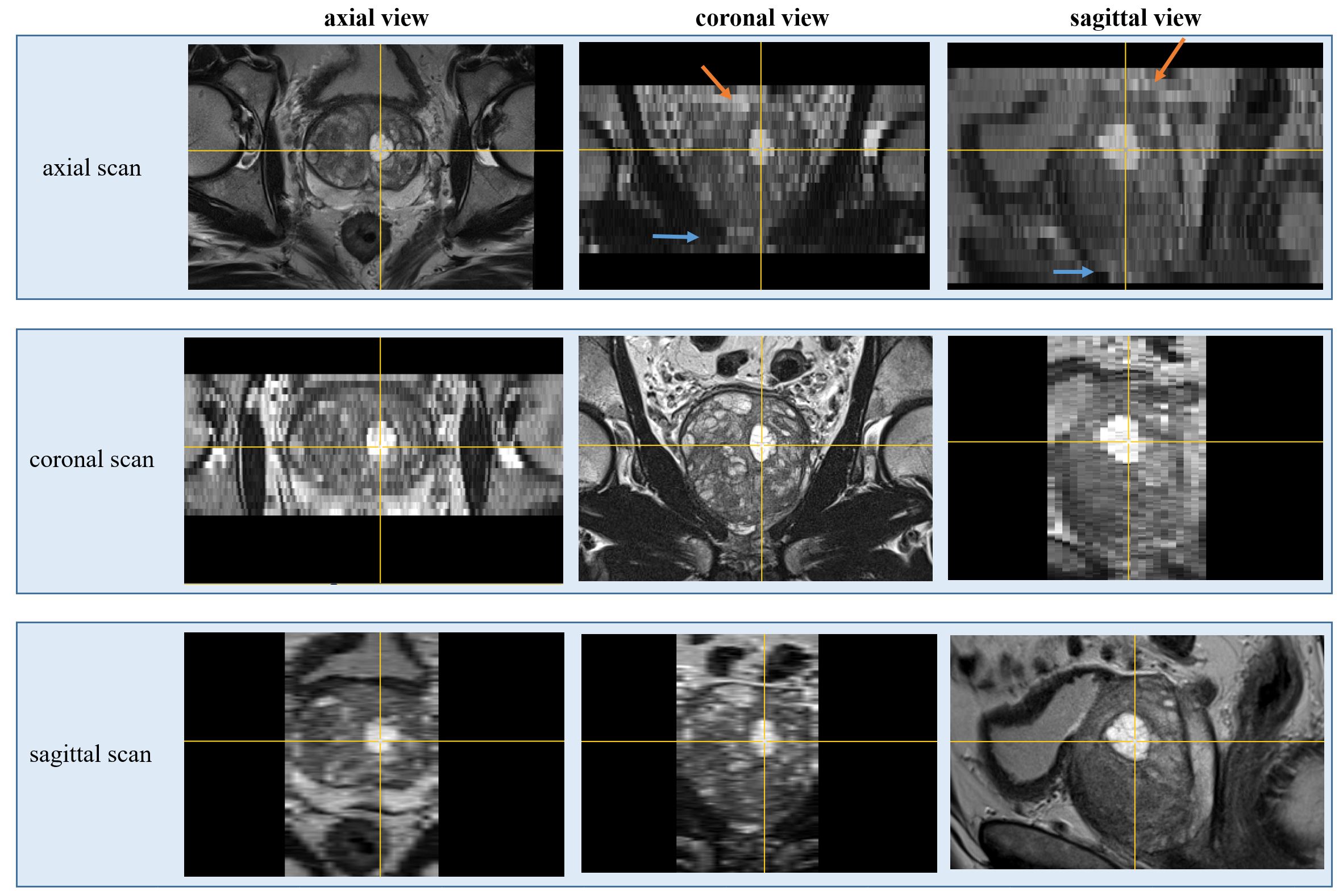}}
%
\caption{Visualization of the independent orthogonal scans of one patient illustrating their anisotropic nature. The first row depicts the axial scan that is normally used for segmentation. As can be seen in the sagittal and coronal view of that axial scan, the apical (blue arrow) and base (orange arrow) region lack clear boundaries of the prostate due to partial volume effect. In the sagittal and coronal scans, the prostate tissue in these regions can be distinguished more clearly from non-prostate tissue.}
\label{fig:motivation}
\end{figure}

The above-mentioned methods use only the axial T2-weighted scan as input, which is suboptimal as MR images acquired in a typical prostate imaging protocol are highly anisotropic (in-plane to out-of-plane resolution ratio of 6-10), see Fig.~\ref{fig:motivation}.
This leads to substantial partial volume artifacts, making it difficult to precisely identify prostate boundaries, especially in the apex and base regions. In addition, segmentations created only on axial volumes suffer from step artifacts due to large slice spacing. However, in prostate cancer imaging protocols as in Weinreb \textit{et al.} \cite{weinreb2016pi}, it is mandatory to acquire at least an additional scan direction (sagittal or coronal) and in multiple clinical routines, all three scan directions are acquired for better interpretation.
These additional scans could be used to improve the prostate segmentation quality, especially in the areas suffering from partial volume effects.

An approach to compute a high-quality prostate mesh was proposed by Shah \textit{et al.}~\cite{Shah2009-ly}, where three masks resulting from manual contouring on axial, coronal, and sagittal MR acquisitions were merged by the means of shape-based interpolation.
Cheng \textit{et al.} introduced a fully automatic segmentation algorithm incorporating multi-planar MR information ~\cite{Cheng2017-zg}.
The algorithm includes an ensemble of three 2D neural networks trained separately on axial, coronal, and sagittal MR scans, respectively. The outputs are fused before a high-resolution prostate segmentation is extracted.
Furthermore, Lozoya \textit{et al.} \cite{Lozoya.10.02.201815.02.2018} assessed the effect of single and dual plane segmentation by training ensembles of 2D CNNs independently on axial and sagittal volumes.
The models process three consecutive image slices (downsampled to a 128$\times$128 resolution) to segment the middle one. The results showed an improvement of 4\% for the dual plane approach.

While these multi-planar approaches show that the exploitation of multi-planar MR images is beneficial for the segmentation quality, they have some limitations.
First, both approaches train independent CNNs per MRI orientation, which prevents the models from learning how to combine the information coming from different orientations.
Second, only 2D neural networks are employed which cannot capture the inherent volumetric information of MRI scans.
Being able to analyze the 3D image context is important for the prostate segmentation as demonstrated by Ye \emph{et al.}\,\cite{L.Yu.2017}, who developed a volumetric ConvNet model that achieved the best performance at the PROMISE12 challenge so far\,\cite{litjens2014evaluation}.
In this work, we address both limitations by presenting a multi-stream 3D CNN architecture that processes simultaneously anisotropic multi-planar MR images to produce a high-resolution prostate segmentation.
This paper builds on our previous work~\cite{Meyer.2018}, where we demonstrated initial results of an~isotropic multi-stream 3D network on a~smaller dataset.\\
Additionally, we evaluate the performance of one, two and three input scan directions on the same dataset. While Lozoya \textit{et al.}~\cite{Lozoya.10.02.201815.02.2018} only include two scan directions, Cheng \textit{et al.}~\cite{Cheng2017-zg} and our previous work~\cite{Meyer.2018} use three planes. All works use different methods and datasets and therefore a thorough investigation of the difference between two and three planes has been impossible so far.

\begin{figure*}[t]
  \centering
  \centerline{\includegraphics[width=13.0cm]{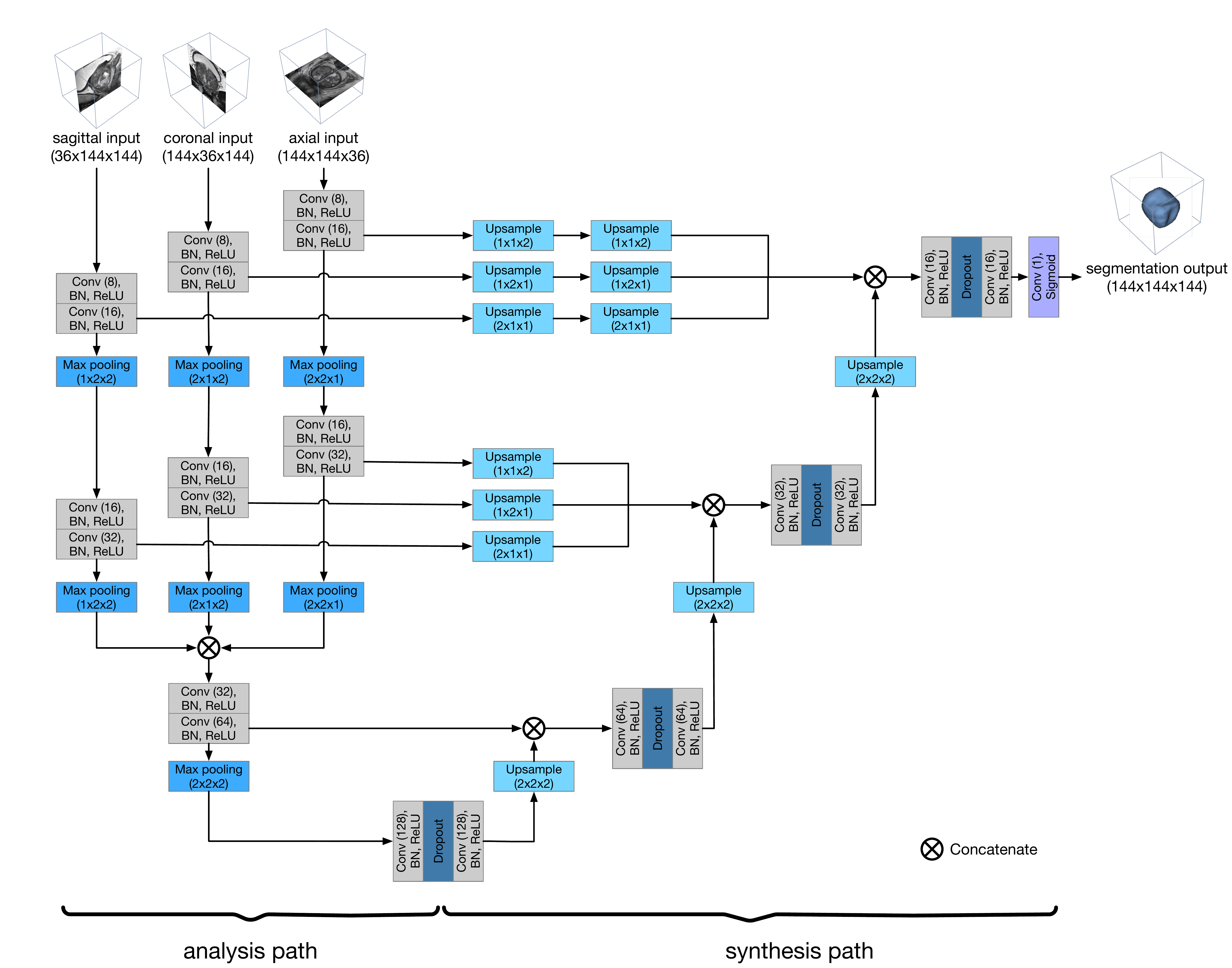}}
  \caption{Triple-planar multi-stream 3D network processing axial, coronal, and sagittal MR volumes. The number in parentheses denotes feature map count (conv layer), pool size (max pooling), and upsampling factor (upsampling). The upsampling is performed either by trilinear upsampling or 3D transposed convolution.}
\label{fig:nets}
\end{figure*}

\subsection{Contribution of this work}

The contribution of this work is two-fold:
\begin{enumerate}
  \item We propose an anisotropic 3D multi-stream CNN architecture and show that it can process multi-planar MR images to produce a high-resolution prostate segmentation. Contrary to our prior work\,\cite{Meyer.2018}, the proposed network design fuses information from anisotropic images alleviating the need for image resampling to isotropic voxel size. Additionally, the proposed architecture is computationally less expensive, which allows for faster inference and more efficient training.

  \item We quantify the influence of information from additional image orientations on segmentation quality by comparing performance of a baseline single-plane model (processing only axial images) with dual-plane (axial + sagittal) and triple-plane (axial + sagittal + coronal) models. To allow a fair comparison of the three approaches, we employ an automatic hyperparameter optimization strategy. We report quantitative results for whole-gland and base, mid and apex regions using image data from two datasets and multiple sites.
\end{enumerate}

Our source code is available on GitHub\footnote{https://github.com/AnnekeMeyer/AnisotropicMultiStreamCNN}, and we published ground truth segmentations that were created as part of this project for a publicly available challenge dataset \cite{segmentations_data}.

In the following section, we describe the proposed architecture for the multi-plane segmentation of the prostate as well as our hyperparameter optimization method. Furthermore, we will give a~description of the datasets used in this work, the training procedure, and the evaluation measures.

\section{Materials and Methods}
\subsection{Multi-Stream Segmentation Network}
With respect to the literature, we can basically define two variants of combining multiple planes for CNNs. The first way is to train three networks separately with each network taking one orthogonal scan as input. The output of the three networks is then fused in a postprocessing step. The alternative is to input all planes in one multi-stream network and and process them simultaneously. We compared the two variants to each other and could not find any significant difference in their performance (see results in Section \ref{sec:ensemble_multi}). Due to its simplicity in deployment, we focus our work on the multi-stream architecture in this project. This has the additional benefit, that we can investigate the influence of additional planes directly, as the ensembling of network outputs has a benefit on performance in general.

Our multi-stream model is a 3D U-Net-like architecture following an encoder/decoder design with four resolution levels~\cite{cciccek20163d}.
The proposed network design is flexible with respect to the number of inputs, which enables information extraction from more than one volume. Fig.~\ref{fig:nets} illustrates a~triple-planar model instance processing axial, coronal, and sagittal acquisitions.
Depending on the desired input configuration (single-plane, dual-plane, or triple-plane), the analysis path has corresponding input specific branches on the first two resolution levels. These branches perform downsampling by max pooling operations with anisotropic pool size (e.g., $2\times2\times1$ for the axial volume) to obtain equally sized outputs, which are concatenated before entering the third resolution level.
Features from the analysis path are passed via long skip connections to the synthesis path. The skip connections contain optionally upsampling layers to bring the feature maps to the corresponding spatial size.
To prevent information bottleneck, we double the filter size in the convolution layers directly followed by max-pooling.
All convolution operations use a $3\times3\times3$ kernel and ReLU nonlinearity.
If batch normalization layers are configured, then they are inserted between the convolution and the nonlinearity.
We employ dropout layers in the synthesis path to avoid overfitting.
The final layer is a $1\times1\times1$ convolution with a sigmoid activation to constrain model output to a $[0, 1]$ range.

\subsection{Hyperparameter Optimization}
Careful tuning of neural network hyperparameters, such as learning rate or regularization strength, is important in getting the best possible model performance. Moreover, hyperparameter optimization\,(HPO) should be performed whenever the architecture or the learned task changes, as a direct transfer of hyperparameter values may lead to a sub-optimal prediction quality. We run HPO to find hyperparameter values yielding the best segmentation performance
for all three architectures (single-, dual-, and triple-plane) independently.
This strategy minimizes the influence of the chosen hyperparameters, yielding a fair comparison among the investigated models.

We employed the HPO strategy that was proposed by Falkner \textit{et al.} in \cite{falkner2018bohb}. The method involves a~combination of Hyperband (HB) with Bayesian optimization (BO) to achieve fast convergence to optimal configurations. HB is an HPO method that evaluates $n$ randomly sampled configurations with a~small budget (e.g., maximal training epoch count), keeps the best half, and doubles their budget \cite{li2016hyperband}. This process is repeated until only one configuration is left. BO builds a probabilistic model based on the already evaluated configurations \cite{shahriari2015taking}. This model is then employed to sample hyperparameter values that should result in better model performance. One iteration of our HPO involves sampling $\tfrac n 2$ configurations from the Bayesian model and another $\tfrac n 2$ by random sampling. The sampled configurations are then evaluated using the HB method.

\subsection{Dataset Description}
We used two datasets for the evaluation of the proposed approaches. The first dataset is an in-house dataset containing 89 axial, sagittal and coronal T2-weighted MR scans acquired on a Philips Achieva 3T imager. In the clinical routine, gland segmentations have been obtained with commercial software (DynaCAD, Philips Invivo) in a semi-supervised manner. As the software only considers the axial T2 volumes, we resampled the segmentations to an isotropic resolution via shape-based interpolation as in Herman \textit{et al.} \cite{herman1992shape}. Subsequently, an expert urologist reviewed and corrected the isotropic segmentations with 3D Slicer \cite{fedorov20123d} by simultaneously considering all three orthogonal scans. \\
The second dataset ProstateX is publicly available through the SPIE-AAPM-NCI Prostate MR Classification Challenge \cite{prostateX_data, lijtens2014prostateX, Clark.2013}, which was designed for predicting the clinical significance of prostate lesions. The dataset comprises multiparametric MRI acquired on two different types of Siemens 3T MR imagers; the MAGNETOM Trio and Skyra. As no reference segmentation of the glands is available in the challenge dataset, we created 66 segmentations for randomly chosen T2-weighted volumes. The segmentations were obtained manually for each scan direction by a medical student, followed by a review and corrections of an expert urologist with 3D Slicer under consideration of all three orthogonal scans.
The final isotropic high-resolution prostate mask is extracted by taking the average of linearly resampled distance transformations of the individual segmentations and thresholding the result at zero (similar to the approach employed by Herman \textit{et al.} \cite{herman1992shape}). The final masks were reviewed by an expert and corrected if necessary using 3D Slicer. These segmentations were published as part of the study to support open research  \cite{segmentations_data, Clark.2013}.

The scans of both datasets were acquired without an endorectal coil. Details on the resolution of the orthogonal scans can be found in Table \ref{tab:scans}. The scans represent prostates with clinical variability such as tumors, cysts, benign prostatic hyperplasia, and scars from previous minimally invasive surgeries.
The alignment of the orthogonal scans was checked visually using 3D Slicer.
In about 10\% of the cases, the scans were misaligned due to, for example, patient or bowel motion.
For these cases, we performed a manual rigid registration of affected images.
Volumes in the ProstateX dataset that did not contain the whole prostate were excluded from this study to have a fair comparison between the approaches. For the in-house dataset, no such cases were found.\\
Methods regarding the segmentation of the prostate glands are often compared to each other in the PROMISE12 challenge \cite{litjens2014evaluation}. As this challenge dataset only consists of axial T2-weighted MR images (see Table \ref{tab:scans}), we were not able to make this comparison in this project.
Instead, we focus on the comparison of different network architectures that are based on the multi-planar input volumes.

\begin{table}[t]
\small
\centering
\caption{Resolution details for Prostate MRI datasets.}

\begin{tabular}{lM{1cm}l}
    \toprule
    \textbf{Dataset}                    & \textbf{Scan} & \textbf{Resolution [mm]} \\
    \midrule
    \multirow{3}{*}{ProstateX} & axial           & {[}0.5-0.6{]} x {[}0.5-0.6{]} x {[}3-5{]}     \\
                                        & sagittal           & 0.56 x 0.56 x {[}3-4{]}                       \\
                                        & coronal           & {[}0.56-0.6{]} x {[}0.56-0.6{]} x {[}3-4.5{]} \\
    \midrule
    \multirow{3}{*}{In-House}  & axial           & 0.5 x 0.5 x 2.75                              \\
                                        & sagittal           & 0.5 x 0.5 x 3.25                              \\
                                        & coronal           & 0.5 x 0.5 x 2.76        \\
    \midrule
    \multirow{3}{*}{PROMISE12}  & axial           & {[}0.27-0.63{]} x {[}0.27-0.63{]} x {[}2.2-3.6{]}                              \\
    & sagittal           & not available                              \\
    & coronal           & not available        \\
    \bottomrule
\end{tabular}
\label{tab:scans}
\end{table}

\subsection{Preprocessing}
For network training and prediction, the three scans are preprocessed by resampling (linear interpolation) them into a common coordinate system. The resulting resolution is 0.5$\times$0.5$\times$2.0mm for axial scans, 0.5$\times$2.0$\times$0.5mm for coronal scans, and 2.0$\times$0.5$\times$0.5mm for the sagittal scans corresponding to their anisotropic acquisition. Next, the images are cropped, such that the resulting volume is the intersection of the three scans. They are further cropped or resized to an in-plane size of 184$\times$184 and an out-of-plane size of 46. As intensity normalization, the gray values are cropped to the 1st and 99th percentiles and afterwards normalized to a range of [0,1].

\subsection{CNN Training}
We set aside randomly chosen 19 test cases for each dataset that were not considered for training. The remaining images were split into four folds for cross-validation. Hence, the folds of the in-house dataset consist of 52 training and 18 validation images each, while the ProstateX fold contains 35 training and 12 validation images.
To augment the training set, random operations such as axial flips, elastic deformations, translations and rotations were used. Unnatural transformations such as top-bottom and front-back flips were not considered. The input images were cropped to a size of 144$\times$144$\times$144~voxels, before being fed to the network.

The objective function of our networks is the negative soft Dice similarity coefficient (DSC)
\[ loss = - \frac{2\sum_{i}^{N}p_{i}g_{i} + \epsilon}{\sum_{i}^{N}p_{i}^{2}+\sum_{i}^{N}g_{i}^{2} + \epsilon},   \]
with \(N\) being the total number of voxels, \(p_i\) and \(g_i\) the predicted and reference voxels, respectively, and $\epsilon$ a small constant to ensure numerical stability.
We ran the training with the Adam optimizer \cite{kingma2014adam} for a maximum of 270 epochs, with an early stop criterium if the validation loss does not improve by at least $\delta = 0.001$ for 100 iterations. The mini-batch size was set to one due to GPU memory capacity (NVIDIA GeForce GTX 1080 Ti).

The prediction was post-processed with a connected components analysis, removing every component except for the largest. We ran the HPO on the concatenation of the first folds from both datasets. For each approach (single, dual and triple-plane), a separate HPO was performed. We optimized the hyperparameters which were empirically found to have substantial influence on model performance: learning rate (range $[10^{-6}, 10^{-3}]$), dropout rate ($0.0, 0.2, 0.4, 0.6$ or $0.8$), upsampling mode\,(tri-linear or transposed convolution), and batch normalization\,(yes or no). The best performing hyperparameters for each approach, selected based on the validation loss, are summarized in Table\,\ref{tab:hyperparameters}. The total numbers of trainable parameters for the single-, dual, and triple-plane of the proposed network architectures are 1.4, 1.6, and 1.7 million, respectively. Thus, the proposed strategies are using similar network capacity.

\begin{table}[t!]
\footnotesize
\def \colwidth {1.5cm}
\centering
\caption{Best performing hyperparameter for each of the investigated network architectures.}
\begin{tabular}{lM{1.5cm}M{\colwidth}M{\colwidth}}
    \toprule
     & \textbf{Single-Plane} & \textbf{Dual-Plane} & \textbf{Triple-Plane} \\
     \midrule
     learning rate & $1.28\times 10^{-4}$ & $1.31\times 10^{-4}$ & $2.99\times 10^{-4}$ \\
     dropout rate & 0.6 & 0.2 & 0.2 \\
     batch normalization & no & no & yes \\
     upsampling mode & tri-linear & transposed convolution & transposed convolution \\
    \bottomrule
\end{tabular}
\label{tab:hyperparameters}
\end{table}

\subsection{Training Scenarios}
We implemented two training scenarios:
\begin{itemize}
  \item \emph{Scenario I} - train one model on merged datasets
  \item \emph{Scenario II} - train separate models for each dataset
\end{itemize}
By comparing models resulting from both scenarios, we can verify whether segmentation quality for a target dataset can benefit from training on multi-site data.
For each scenario, four-fold cross-validation was performed.

\subsection{Evaluation Measures}
We evaluated the investigated models with the following measures that were also used in the PROMISE12 challenge\,\cite{litjens2014evaluation}: Dice similarity coefficient (DSC) as well as the average boundary distances (ABD) and the 95th percentile Hausdorff-Distance (95-HD) between surface points of both volumes.

The Dice similarity coefficient is defined as

\begin{equation}
\text{DSC}(X, Y) = \frac{2 \, |X \cap Y|}{|X|+|Y|}
\end{equation}

with $X$ being the predicted and $Y$ being the ground truth voxels. The average boundary distance is defined as:

\begin{equation}
    \begin{split}
        \text{ABD}(X_{S}, Y_{S}) = \frac{1}{|X_S|+|Y_S|} ( &\sum_{x \in {X_S}} \min_{y \in Y_S}\text{ED}(x,y)\\
        +  &\sum_{y \in {Y_S}} \min_{x \in X_S}\text{ED}(y,x))
    \end{split}
\end{equation}

where $X_S$ and $Y_S$ are the sets of surface points of the predicted and ground truth segmentation. $\text{ED}$ is the Euclidean distance operator. The Hausdorff distance is defined as

\begin{equation}
    \begin{split}
    \text{HD}(X_S, Y_S) & = \max \left(\text{HD}'(X_S, Y_S), \text{HD}'(Y_S, X_S)\right)\\
    \text{with HD}'(X_S, Y_S) & = \max_{x \in X_S} ( \min_{y \in Y_S}\text{ED}(x,y)).
    \end{split}
\end{equation}

As done in \cite{litjens2014evaluation}, we used the 95th percentile for implementation of HD (the so-called 95-HD), as this measure is more often used, leveraging comparability with previous works.

All evaluation measures are computed in 3D each for the whole gland, apex, base, and mid-gland regions. Each region corresponds to ca. one-third of the prostate and was partitioned in a slice-based manner with regards to the manual reference segmentation.

\section{Results and Discussion}

\begin{table*}[t!]
\centering
\footnotesize
\setlength\tabcolsep{3pt}
\caption{Evaluation measures for scenario I (training on merged datasets) averaged across all folds.
Asterisks mark significantly better results when compared to the single-plane model.}
\begin{tabular}{ll|lll|lll|lll}
\toprule
                       &       & \multicolumn{3}{c|}{Merged Datasets} & \multicolumn{3}{c|}{ProstateX}  & \multicolumn{3}{c}{In-House}             \\
                       &       & Single & Dual           & \multicolumn{1}{l|}{Triple}   & Single & Dual           & \multicolumn{1}{l|}{Triple} & Single         & Dual           & Triple  \\
\midrule
\multirow{4}{*}{DSC}   & Whole & $0.927$  & $\textbf{0.933}^{**}$ & $0.931^{*}$   & $0.917$  & $\textbf{0.925}^{*}$ & $0.922$        & $0.936$ & $\textbf{0.941}^{*}$ & $0.939$  \\
                       & Apex  & $0.888$  & $\textbf{0.901}^{*}$ & $0.896$   & $0.854$  & $\textbf{0.880}^{***}$ & $0.872$        & $\textbf{0.922}$ & $0.921$ & $0.920$  \\
                      & Mid   & $0.956$  & $\textbf{0.958}^{*}$ & $0.954$   & $\textbf{0.957}$  & $0.956$ &  $0.950$          & $0.955$ & $\textbf{0.960}^{*}$ & $0.959$  \\
                       & Base  & $0.898$  & $0.904^{**}$ & $\textbf{0.906}^{*}$   & $0.884$  & $0.890^{*}$ & $\textbf{0.893}$       & $0.912$ & $\textbf{0.919}$ & $0.918$  \\
\midrule
\multirow{4}{*}{ABD[mm]}   & Whole & $0.901$  & $\textbf{0.841}^{**}$ & $0.877$ & $1.088$  & $\textbf{1.019}^{*}$ & $1.048$  & $0.714$  & $\textbf{0.664}^{*}$ & $0.705$  \\
                       & Apex  & $0.990$  & $\textbf{0.863}^{*}$ & $0.916$   & $1.343$ & $\textbf{1.084}^{***}$ & $1.160$         & $\textbf{0.637}$  & $0.643$ & $0.672$  \\
                       & Mid   & $\textbf{0.762}$  & $0.779^{*}$ & $0.827$   & $\textbf{0.797}$  & $0.918$ & $0.971$        & $0.727$ & $\textbf{0.640}^{**}$ & $0.684$  \\
                       & Base  & $1.007$  & $\textbf{0.946}^{*}$ & $0.947^{*}$   & $1.230$  & $1.176$ & $\textbf{1.143}$        & $0.783$ & $\textbf{0.715}$ & $0.751$  \\
\midrule
\multirow{4}{*}{95-HD[mm]} & Whole & $3.101$  & $\textbf{3.005}^{*}$ & $3.072$   & $3.916$  & $3.927$ & $\textbf{3.721}$  & $2.286$ & $\textbf{2.083}^{*}$ & $2.422$  \\
                       & Apex  & $2.992$  & $\textbf{2.651}^{*}$ & $2.810$   & $4.017$  & $\textbf{3.288}^{***}$ & $3.520^{*}$        & $\textbf{1.967}$ & $2.015$ & $2.100$  \\
                       & Mid   & $\textbf{2.483}$  & $2.687^{**}$ & $2.740$   &  $\textbf{2.754}$  & $3.439$ & $3.212$        & $2.213$ & $\textbf{1.935}^{*}$ & $2.269$  \\
                       & Base  & $3.097$  & $2.932^{*}$ & $\textbf{2.899}$   & $3.670$  & $3.706$ &  $\textbf{3.478}$       & $2.524$ & $\textbf{2.158}^{*}$ & $2.321$  \\
\bottomrule
\multicolumn{11}{l}{Best results are marked bold. ${}^*p<0.05$, ${}^{**}p<0.01$, ${}^{***}p<0.001$.}
\end{tabular}
\label{tab:results}
\end{table*}

\begin{table*}[t]
\centering
\footnotesize
\caption{
Evaluation measures for scenario II (models are trained and evaluated on each dataset individually) averaged across all folds.
Asterisks mark significantly better results when compared to the single-plane model.
}
\begin{tabular}{ll|lll|lll}
\toprule
                       &       & \multicolumn{3}{c|}{ProstateX}  & \multicolumn{3}{c}{In-House}             \\
                       &       & Single & Dual           & \multicolumn{1}{l|}{Triple} & Single         & Dual           & Triple  \\
\midrule
\multirow{4}{*}{DSC}   & Whole & 0.919  & 0.923$^{*}$ & \textbf{0.926}          & 0.927          & \textbf{0.939} & \textbf{0.939}$^{*}$          \\
                       & Apex  & 0.865  & 0.873 & \textbf{0.875}            & 0.917          & 0.919          & \textbf{0.920}  \\
                       & Mid   & \textbf{0.956}  & 0.952 & 0.953          & 0.946          & \textbf{0.960}          & 0.959  \\
                       & Base  & 0.886  & 0.896 & \textbf{0.900}$^{*}$          & 0.897         & 0.914          & \textbf{0.915}  \\
\midrule
\multirow{4}{*}{ABD[mm]}   & Whole & 1.056  & 1.014$^{*}$ & \textbf{0.994}   & 0.793 & 0.704          & \textbf{0.680}           \\
                       & Apex  & 1.228  & 1.144 & \textbf{1.118}          & 0.673          & 0.677          & \textbf{0.662}  \\
                       & Mid   & \textbf{0.808}  & 0.906 & 0.910          & 0.810          & \textbf{0.652} & 0.658           \\
                       & Base  & 1.207  & 1.094$^{*}$ & \textbf{1.065}$^{*}$          & 0.904          & 0.785          & \textbf{0.729}  \\
\midrule
\multirow{4}{*}{95-HD[mm]} & Whole & 3.731  & \textbf{3.600} & 3.666  & 2.604 & 2.405          & \textbf{2.155}           \\
                       & Apex  & 3.573  & \textbf{3.393} & 3.413       & 2.076          & 2.052 & \textbf{1.999}           \\
                       & Mid   & \textbf{2.726}  & 3.054 & 3.375          & 2.482       & \textbf{2.004}  & 2.016           \\
                       & Base  & 3.718  & \textbf{3.347} & 3.456          & 2.920          & 2.676          & \textbf{2.096}$^{**}$ \\
\bottomrule
\multicolumn{8}{l}{Best results are marked bold. ${}^*p<0.05$, ${}^{**}p<0.01$, ${}^{***}p<0.001$.}
\end{tabular}
\label{tab:results_sep}
\end{table*}

We report quantitative results (averaged across folds) of both scenarios in Table \ref{tab:results} and Table \ref{tab:results_sep}, respectively. Each approach was subject to four-fold cross-validation, and the performance of the resulting models was evaluated on left-out test cases. We applied the Wilcoxon signed-rank test to obtain the statistical significance of quantitative differences between single and dual or triple plane approaches. The rationale against a standard Student t-Test is that we cannot assume Gaussianity for the distribution of the result quality.

In general, the additional scans used by the dual- and triple-plane models improved the segmentation quality when compared with the single-plane model. In the following, we present a more detailed result analysis for both considered scenarios as well as comparison with the inter-rater variability.

\subsection{Scenario I}

\begin{figure*}
\centering
\subfloat[DSC]{%
  \includegraphics[width=0.6\textwidth]{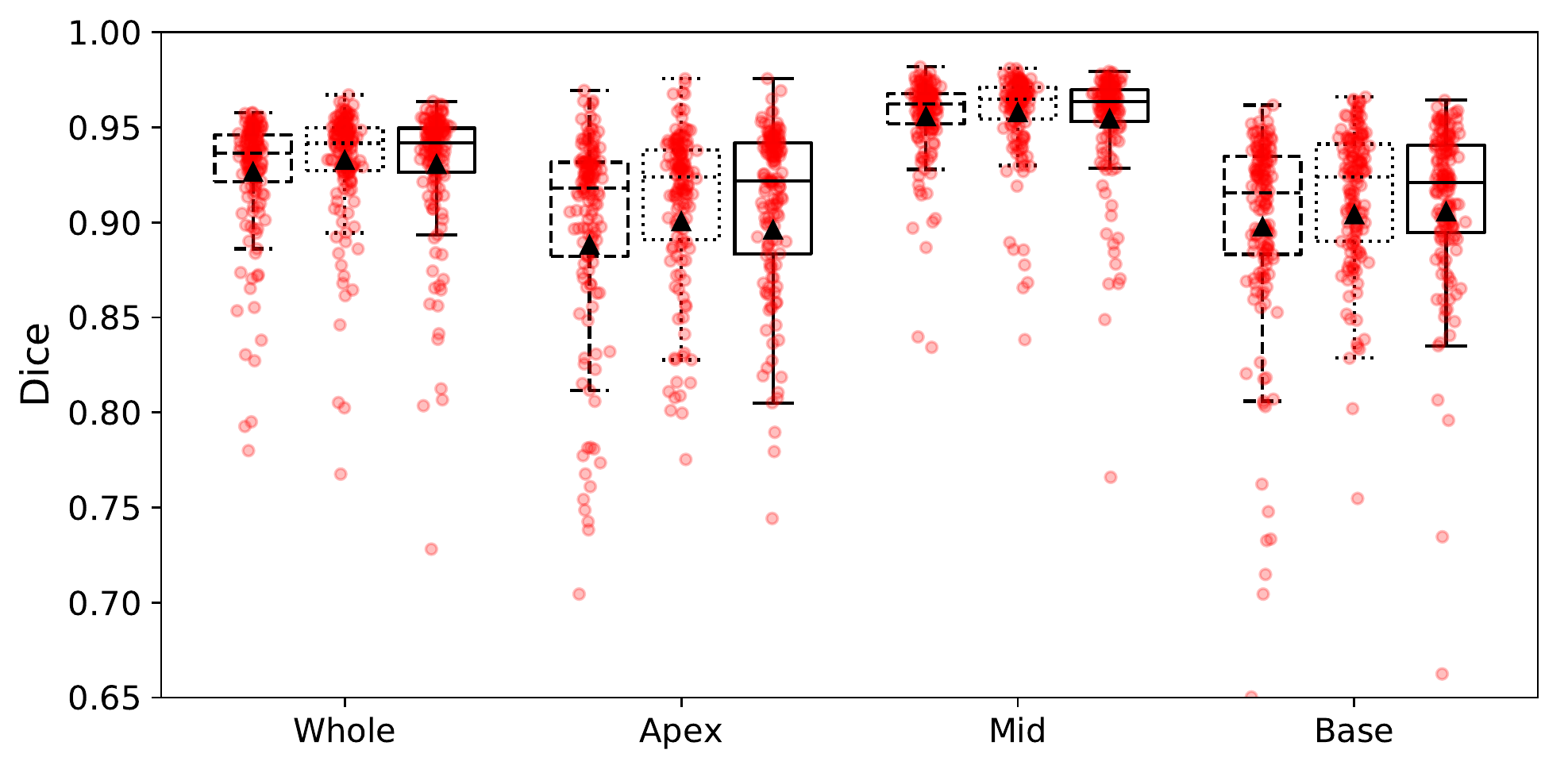}%
  \label{DSC}
}\\
\subfloat[ABD]{%
  \includegraphics[width=0.6\textwidth]{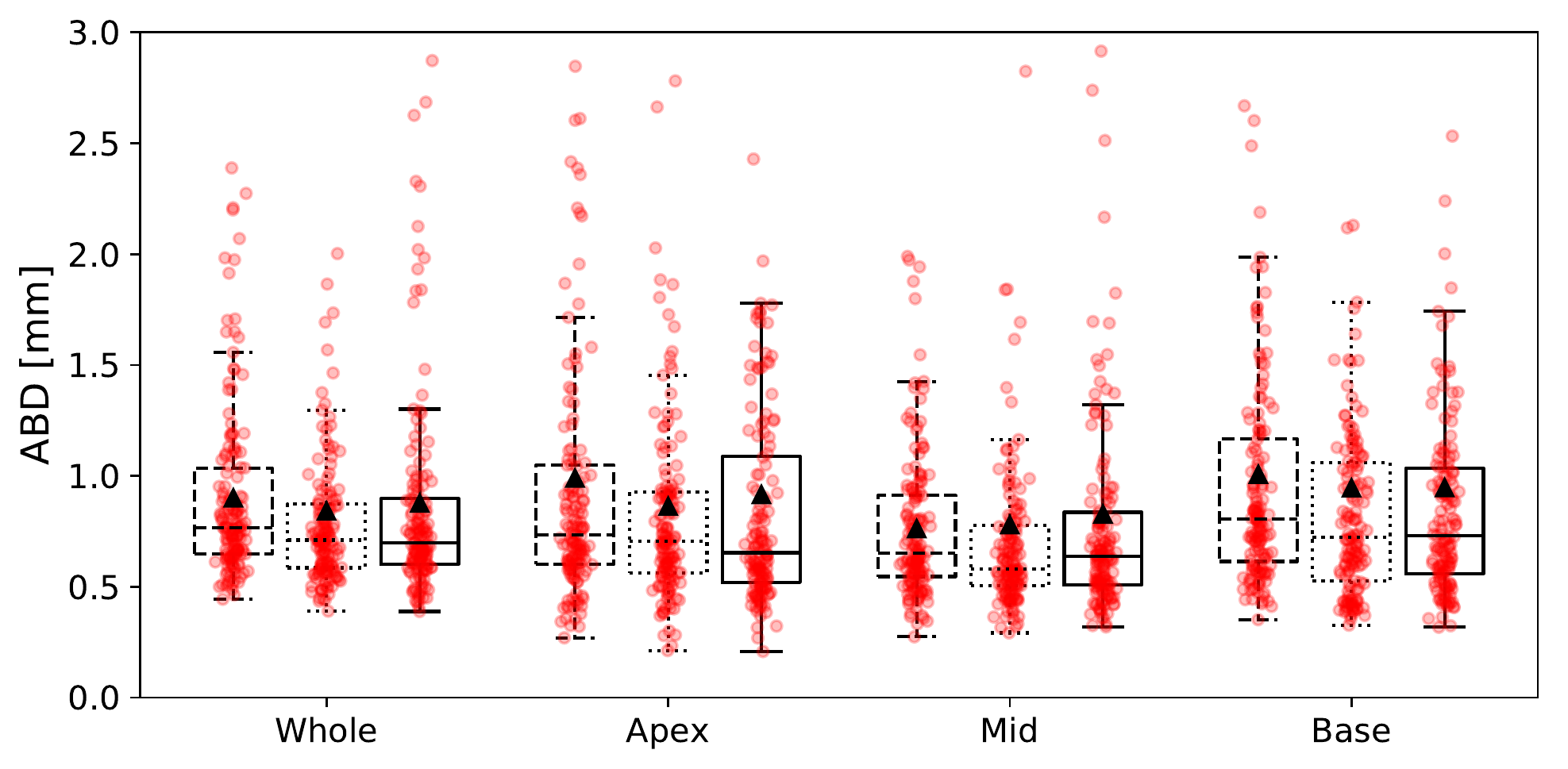}%
  \label{ABD}
}\\
\subfloat[95-HD]{%
  \includegraphics[width=0.6\textwidth]{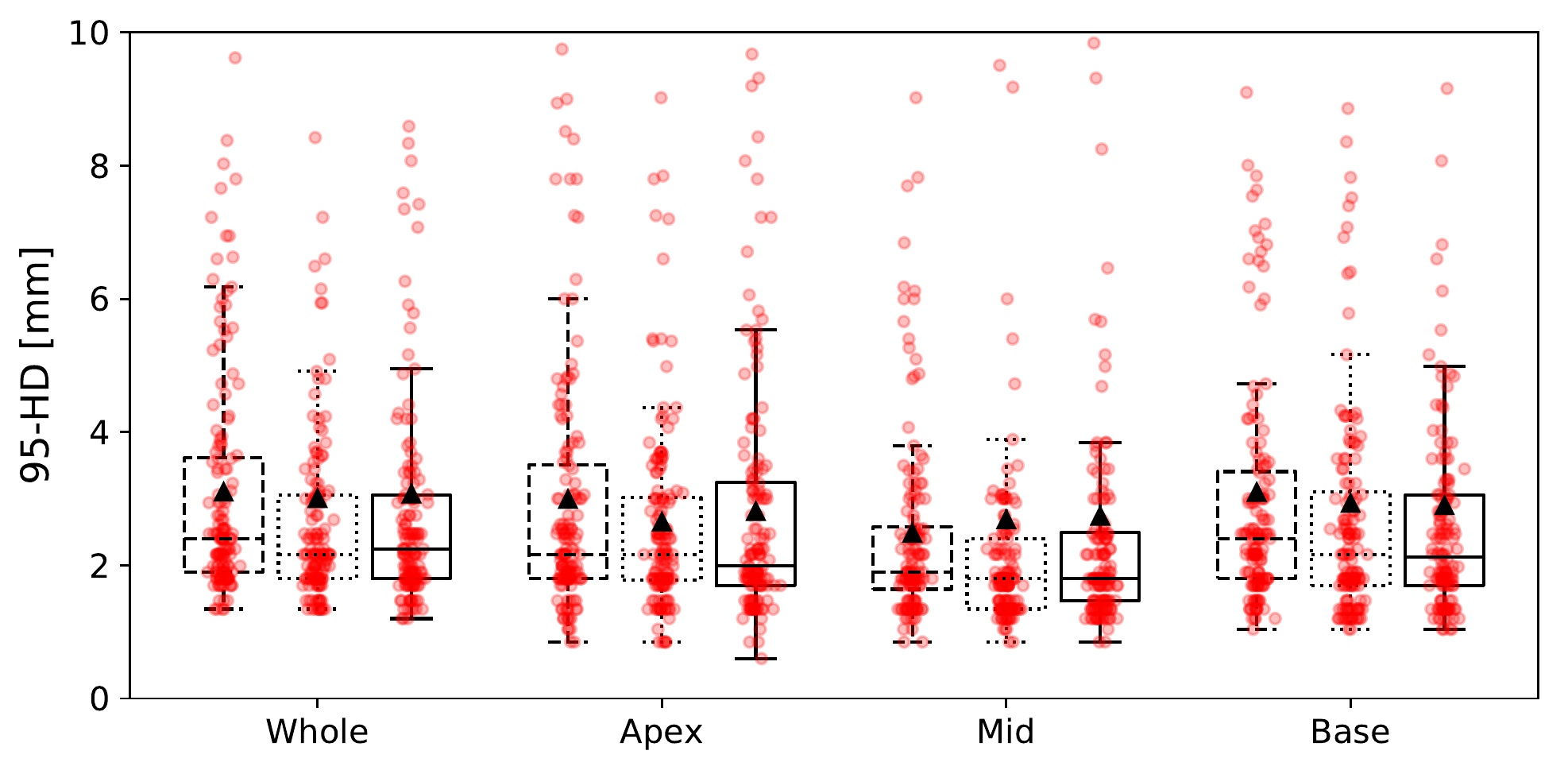}%
  \label{95-HD}
}
\caption{Boxplots showing (a)\,DSC, (b)\,ABD, and (c)\,95-HD for the whole gland and its subregions for single-\,(dashed), dual-\,(dotted), and triple-plane\,(solid) models. Models were trained on merged datasets (scenario I).}
\label{fig:boxplots_results_scenarioI}
\end{figure*}

In training scenario I (training on merged datasets), the dual-plane approach that incorporates axial and sagittal volumes, works significantly better ($p<0.05$) than the single-plane approach on both datasets and every region of the prostate with regards to every evaluation measure. The dual-plane method achieved an average DSC of 0.933 for the whole gland (vs. 0.927 for single-plane), 0.901 (vs. 0.888) in the apex and 0.958 (vs. 0.956) and 0.904 (vs. 0.898) for mid-gland and base, respectively. It has to be noted that the ABD and 95-HD for the mid-region are worse for dual-plane than single-plane, but the boxplots in Fig. \ref{fig:boxplots_results_scenarioI} indicate that the dual-plane model performs better when the median is considered.
The triple-plane model performed significantly better ($p<0.05$) than the single-plane model regarding the DSC of the whole prostate as well as of the base region.
Regarding distance-based measures, only the ABD of the base region was significant ($p<0.05$).

\begin{figure*}[htp]
\centering
\subfloat[Simple case where all approaches perform about equally well.]{%
  \includegraphics[width=\textwidth]{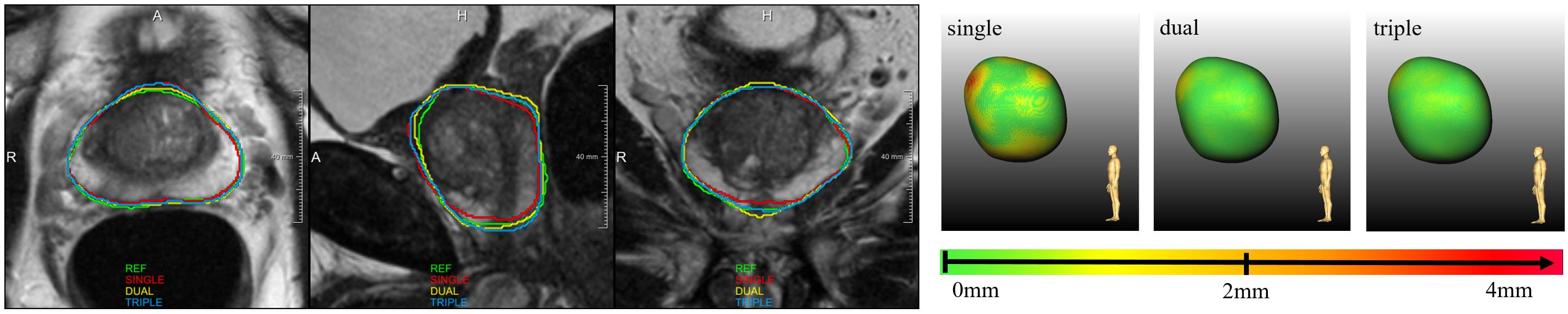}%
  \label{example1}
}

\subfloat[Challenging case where dual/triple plane approaches are necessary. When considering only the axial plane, we yield overestimation in the base region.]{%
  \includegraphics[width=\textwidth]{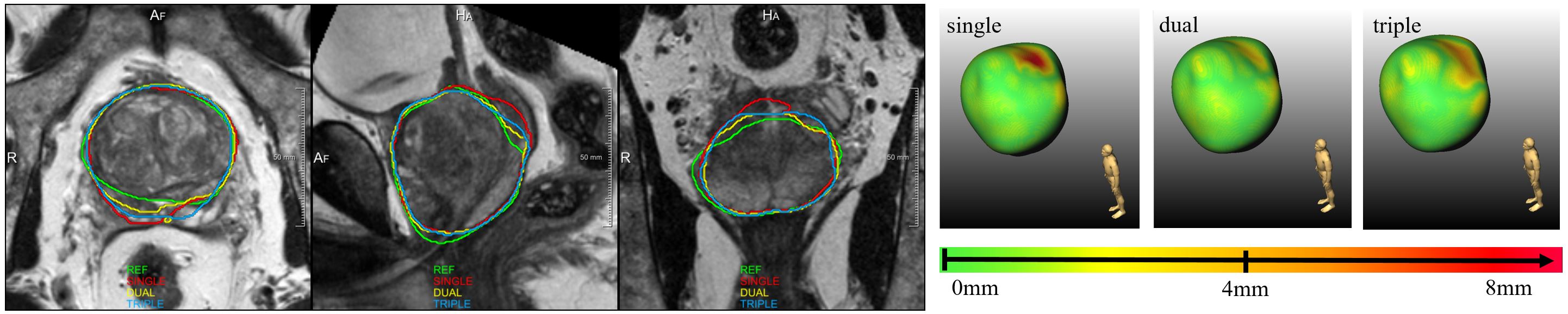}%
  \label{example3}
}

\subfloat[Challenging case where dual/triple plane approaches are necessary. Segmentation in apical region of the prostate is improved.]{%
  \includegraphics[width=\textwidth]{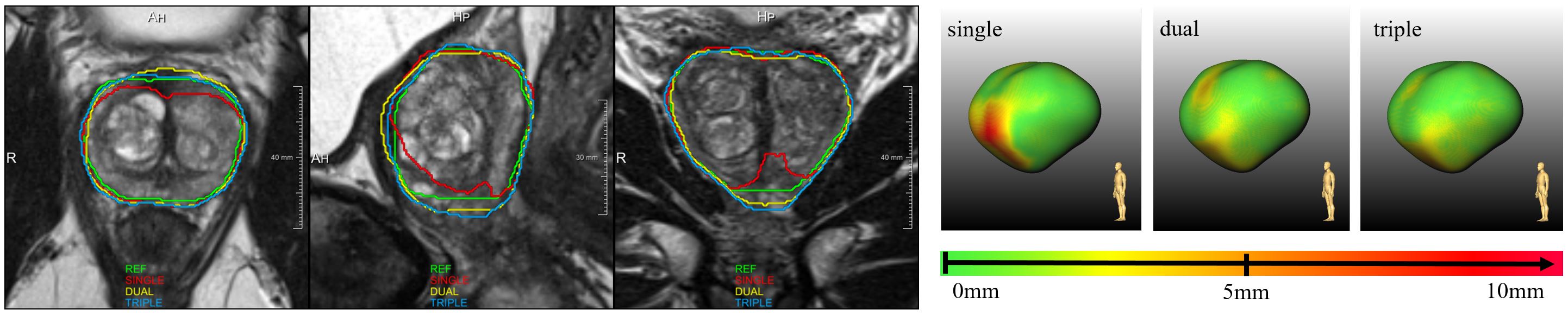}%
  \label{example4}
}

\subfloat[Challenging case, where all approaches fail, presumably due to strong heterogeneity in the prostate gland.]{%
  \includegraphics[width=\textwidth]{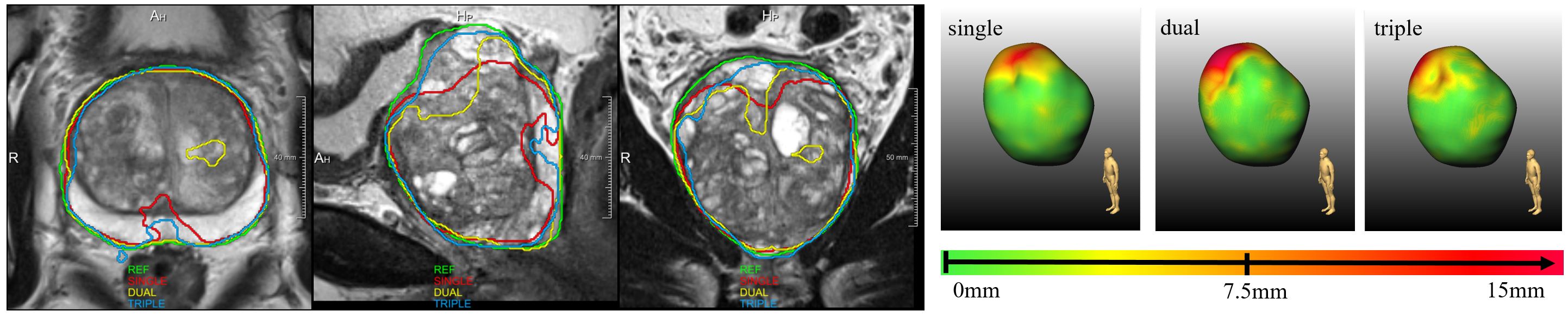}%
  \label{example5}
}

\caption{Four examples with different characteristics. On the left, segmentations in the image plane are depicted. Left column is the axial view, central column is sagittal view, and right column depicts the coronal view. On the right the surface distance between ground truth and prediction are shown for each approach.}
\label{fig:visual_examples}
\end{figure*}

Overall, the difference in performance between dual and triple-plane is less than between single-plane and triple- or dual-plane for training scenario\,I. 
Examples for the above-described segmentation quality differences are depicted in Fig.\,\ref{fig:visual_examples}.

\subsection{Scenario II}

For scenario\,II, we can find less significant differences between the different approaches (see Tab.\,\ref{tab:results_sep}). This may be caused by the fact that less training data was available for each experiment.
Opposed to scenario\,I (Table \ref{tab:results}), where the dual-plane approach achieved the best performance for the evaluation measures in general, the triple-plane approach generally performs better in scenario II than dual-plane for each region and evaluation measure.

\subsection{Scenario I vs. II}
\begin{figure*}
  \centering
  \centerline{\includegraphics[width=1.2\textwidth]{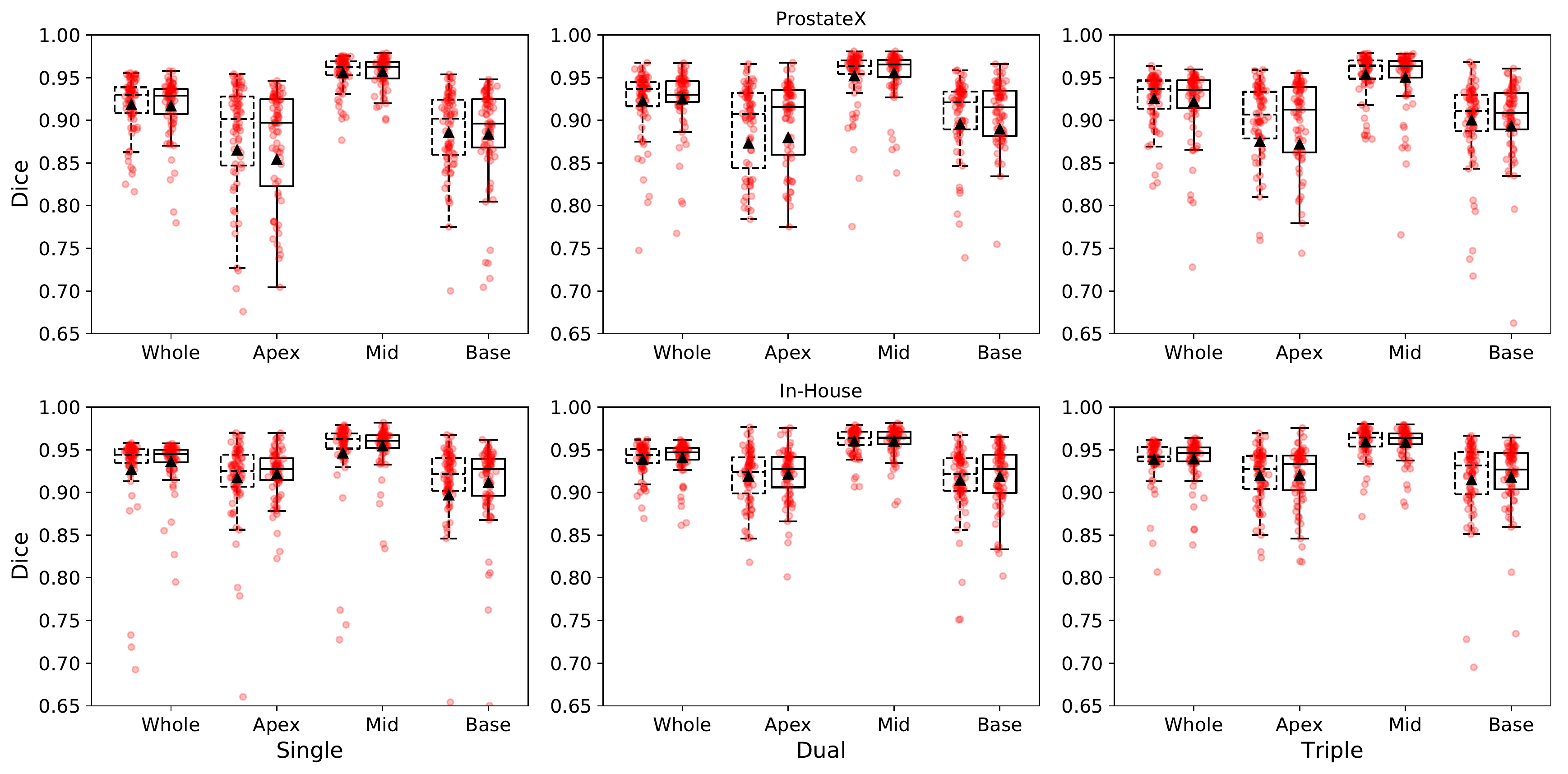}}
\caption{Boxplots comparing Dice similarity coefficients for the whole gland and its subregions for models trained on only one dataset (scenario II, dashed) and merged datasets (scenario I, solid). Results are accumulated from all folds. The quality differences between scenario I and II are not substantial, yet, a slight improvement for the whole gland and most regions for models resulting from scenario I can be observed.}
\label{fig:compare_scenarios}
\end{figure*}

In general, the differences in performance between scenario\,II (training on individual datasets) and scenario\,I (training on merged datasets) were not substantial.
However, we can see a slight improvement in the boxplots in Fig.\,\ref{fig:compare_scenarios} for the whole gland and most regions when models are trained on merged datasets.

We observed that the quantitative evaluation measures in both scenarios are considerably better for the in-house datasets than for the ProstateX data. We assume that the reason for these results is two-fold: Firstly, the number of cases in the datasets are not balanced: the in-house dataset had almost 50\% more cases available for training (n=70) than the ProstateX dataset (n=47). Secondly, the reference annotations were created with different methods: while the annotations for the ProstateX dataset were created entirely manually, the in-house dataset was segmented semi-automatically in the first stage and later refined manually. Even when experts review and correct the semi-automatically generated segmentations, there may still be a potential bias towards the semi-automatic segmentations, which could result in more consistent segmentations than with manual delineation. One might also argue that the image quality is another factor for performance quality. However, we could not confirm this visually.

Another observation we made is that, on the one hand, the triple-plane approach performs better than dual-plane if models are trained on separate datasets (scenario II). On the other hand, the dual-plane approach is more often significantly better than single-plane as the triple–plane approach is when trained on merged datasets (scenario I). Thus, dual-plane seems to be more robust to variations in the training data if multiple data sources are used. However, the quantitative differences between dual- and triple-plane in both training scenarios are not statistically significant.

\subsection{Inter-Rater Variability}
To put our automatic segmentation results into perspective, we were interested to see in what range the inter-observer variability of prostate segmentation is (see Table \ref{tab:results_inter_rater}). In the literature, second observer segmentation evaluation has been investigated within the scope of the PROMISE12 challenge \cite{litjens2014evaluation}. The authors report a mean DSC of 0.90 between two expert segmentations for the whole gland and 0.80 and 0.86 for the apex and base, respectively. For the whole gland, they report an inter-rater variability of 5.64\,mm for 95-HD.

We carried out a similar study as part of another project where we asked two urologists to outline the glandular structures in the axial scans of 20 cases from the ProstateX challenge \cite{meyer2019towards}. It has to be noted that those cases do not cover the test cases of this work. Nevertheless, we can still get a notion of how much two expert segmentations can vary. The inter-rater DSC for the whole gland, apex and base for these 20 cases were 0.93, 0.90 and 0.89, respectively. The 95-HD was 3.15\,mm for the whole gland, which corresponds approximately to the thickness of one slice. Comparing these results to the overall DSC of 0.93 for the dual- and triple-planar model, we are clearly in the range of inter-rater variability. However, individual cases, as shown in Fig. \ref{example5}, still indicate that automatic segmentations need to be further improved in the future.

\subsection{Multi-Stream vs. Ensemble}
\label{sec:ensemble_multi}
We compared our triple-plane architecture processing all orthogonal images simultaneously, which directly outputs a prostate segmentation, with an ensemble approach from the literature\,\cite{Cheng2017-zg, Lozoya.10.02.201815.02.2018}. In the ensemble approach three independent 3D models are trained for each image orientation, which outputs are combined in a post-processing step. We trained three single-plane models and used majority vote to compute the final segmentation.
The experiment was performed on the ProstateX dataset and results were averaged across 4 folds.
The results are listed in Table \ref{tab:results_ens_multi_stream}. No significant differences were found between the two methods for any region and evaluation measure (Wilcoxon signed-rank test). The results are also in line with the outcome of our study that the input of multiple planes improves over a single-plane input.\\
Although no differences were found, we think that the multi-stream approach is superior to the ensemble because it requires less parameters (factor of 2.7) and therefore is easier to deploy in production.
Moreover, using common decoder for all image orientations (as in multi-stream architecture) can be seen as a regularizer, which can help in minimising the generalization error on other datasets/tasks.
For the ensemble we also evaluated output combination using shape-based interpolation, but it worked worse than majority vote.

\begin{table}
\footnotesize
\centering
\caption{Evaluation measures for inter-rater variability}
\label{tab:results_inter_rater}
\begin{tabular}{ll|c|c}
\toprule
                       &       & \multicolumn{2}{c}{Inter-Observer}   \\
                       &       & PROMISE12 (n=30) & ProstateX (n=20)  \\
\hline
\multirow{4}{*}{DSC}   & Whole & 0.90             & 0.93              \\
                       & Apex  & 0.80             & 0.90              \\
                       & Mid   & n.a.             & 0.96              \\
                       & Base  & 0.86             & 0.89              \\
\hline
\multirow{4}{*}{ABD}   & Whole & 1.82             & 0.66              \\
                       & Apex  & 2.55             & 0.63              \\
                       & Mid   & n.a              & 0.49              \\
                       & Base  & 2.21             & 0.86              \\
\hline
\multirow{4}{*}{95-HD} & Whole & 5.64             & 3.15              \\
                       & Apex  & 6.36             & 2.84              \\
                       & Mid   & n.a              & 2.02              \\
                       & Base  & 6.28             & 3.56             \\
\bottomrule
\end{tabular}
\end{table}

\begin{table}
\footnotesize
\centering
\caption{Comparison of two methods (ensemble and triple-plane) for generating segmentations from tri-planar input. No significant differences were found.}
\label{tab:results_ens_multi_stream}
\begin{tabular}{ll|ll}
\toprule
                           &       & \multicolumn{1}{c}{ensemble}        & \multicolumn{1}{c}{triple-plane}            \\
\midrule
\multirow{4}{*}{DSC}       & Whole & $0.926 \pm 0.03$   & $0.926 \pm 0.03$   \\
                           & Apex  & $0.871 \pm 0.12$  &  $0.875 \pm 0.12$   \\
                           & Mid   & $0.955 \pm 0.02$   & $0.953 \pm 0.03$   \\
                           & Base  & $0.901 \pm 0.04$   & $0.900 \pm 0.04$   \\
\midrule
\multirow{4}{*}{ABD[mm]}   & Whole & $0.947 \pm 0.36$   &  $0.994 \pm 0.46$   \\
                           & Apex  & $0.871 \pm 0.12$  & $0.875 \pm 0.12$     \\
                           & Mid   & $0.789 \pm 0.30$ & $0.910 \pm 0.63$      \\
                           & Base  & $1.028 \pm 0.56$ & $1.065 \pm 0.60$     \\
\midrule
\multirow{4}{*}{95-HD[mm]} & Whole & $3.10 \pm 1.38$ & $3.666 \pm 2.23$     \\
                           & Apex  & $3.13 \pm 1.72$ & $3.413 \pm 1.78$     \\
                           & Mid   & $2.29 \pm 0.99$ & $3.375 \pm 3.26$     \\
                           & Base  & $3.01 \pm 1.70$ & $3.456 \pm 2.17$     \\
\bottomrule
\end{tabular}
\end{table}

\section{Conclusion and Future Work}
We proposed an anisotropic 3D multi-stream segmentation CNN that allows incorporating of different numbers of orthogonal input volumes. The objective of our work was to determine whether segmentation performance could be increased by the incorporation of sagittal and coronal volumes. To allow for a fair comparison between single-, dual- and triple-plane approaches, we included an automatic hyperparameter optimization strategy.

The most important finding of this work is that the use of multi-planar strategies significantly improves segmentation performance compared to using only axial volumes in almost all cases. The quantitative differences between the three proposed approaches may not be large, but depending on the clinical application, the improved accuracy can be critical for the conservation of structures like external sphincter, bladder, or seminal vesicles.
The clinical utility of the multi-planar approaches would be addressed in future work.
Whether to prefer using the dual- or triple-plane variant could not be answered unequivocally. However, the dual-plane approach seems to be a good trade-off between computational cost and segmentation quality.

Future work will include an automatic registration among the orthogonal scans to compensate for potential transformations between them.
This may lead to an increased performance of the multi-planar approaches, as the manual registration may not compensate for all motion artifacts and may be less precise than an automatic method.
Another field of future research will be the detailed investigation of the multi-stream network architecture. For example, the location where the encoders are merged could be further examined.

Our results quality is comparable to the inter-rater variability. However, as mentioned above, some negative outliers would have never been produced by any medical experts. Hence, future work should also investigate how those outliers could be automatically detected and how much correction time would be required to achieve clinically acceptable segmentations.
Furthermore, it would be interesting to apply our multi-stream architecture to other clinical use cases, where multi-planar imaging is acquired (e.g., cardiac MRI).

\section*{Conflict of Interest}
This work has been funded by the EU and the federal state of Saxony-Anhalt, Germany under grant number ZS/2016/08/80388. Co-Funding was provided by Fraunhofer-Society. The Titan Xp used for this research was donated by the NVIDIA Corporation. Data used in this research were obtained from The Cancer Imaging Archive (TCIA) sponsored by the SPIE, NCI/NIH, AAPM, and Radboud University.

\bibliographystyle{elsarticle-num}

\end{document}